\documentclass[superscriptaddress, reprint, amsmath, amssymb, aps, pra, floatfix]{revtex4-2}

\usepackage[utf8]{inputenc}
\usepackage{graphicx}
\usepackage{dcolumn}
\usepackage{bm}
\usepackage[normalem]{ulem} 
\usepackage{mathrsfs}
\usepackage{physics}
\usepackage{xcolor}
\usepackage{amsmath, amssymb, mathtools, amsthm}
\usepackage[colorlinks=true]{hyperref}
\usepackage{comment} 
\usepackage{orcidlink}

\usepackage{xr}

\graphicspath{{Figures/}} 

\definecolor{ve}{RGB}{0,161,22}

\begin{abstract}
Continuous measurement is commonly associated with decoherence and the loss of symmetry. Here, we show that weak continuous measurements can instead act selectively: depending on the measured observable, they may either preserve or remove a dynamical relation between initially mirrored states. Using tunneling between the $\ket{+1}$ and $\ket{-1}$ states of a spin-1 system, we demonstrate that measurements of $S_x$ and $S_y$ break the symmetry of the tunneling dynamics in the presence of a longitudinal bias, whereas measurement of $S_z$ leaves it intact. We formulate this behavior within the Lindblad framework, support it with numerical simulations, and reproduce the dissipative dynamics in a two-qubit quantum circuit using Trotterized evolution and stochastic collective rotations. Our results establish continuous measurement as a selective tool for controlling dynamical symmetry in open quantum systems.

\end{abstract}

\begin{document}
\author{Polina Kofman\orcidlink{0000-0002-7623-3463}}
\email{polinaokofman@gmail.com}
\affiliation{Departamento de Fisica, Instituto Superior Tecnico, Universidade de Lisboa, Av. Rovisco Pais, 1049-001 Lisboa, Portugal}
\affiliation{B. Verkin ILTPE of NASU, 47 Nauky Ave., Kharkiv, 61103, Ukraine}

\author{Elvira Bilokon\orcidlink{0009-0007-8296-2906}}
\email{ebilokon@tulane.edu}
\affiliation{Department of Physics and Engineering Physics, Tulane University, New Orleans, Louisiana 70118, United States}
\affiliation{Akhiezer Institute for Theoretical Physics, NSC KIPT, Akademichna 1, 61108 Kharkiv, Ukraine}

\author{Valeriia Bilokon\orcidlink{0009-0001-1891-0171}}
\email{vbilokon@tulane.edu}
\affiliation{Department of Physics and Engineering Physics, Tulane University, New Orleans, Louisiana 70118, United States}
\affiliation{Akhiezer Institute for Theoretical Physics, NSC KIPT, Akademichna 1, 61108 Kharkiv, Ukraine}

\author{Denys I. Bondar \orcidlink{0000-0002-3626-4804}}
\email{dbondar@tulane.edu}
\affiliation{Department of Physics and Engineering Physics, Tulane University, New Orleans, Louisiana 70118, United States}

\title{Measurement-Selective Dynamical Symmetry Breaking}

\date{\today}

\maketitle

\section{Introduction}
Symmetry is one of the central organizing principles of physics. It determines which quantities are conserved, constrains the dynamics of quantum systems, and plays a key role in phase transitions and quantum information. In isolated systems, a symmetry is conventionally associated with an operator that commutes with the Hamiltonian and, through Noether's theorem, gives rise to a conserved quantity. Realistic quantum systems, however, are never perfectly isolated, and coupling to an environment can modify or destroy such conservation laws. This has motivated a broader formulation of symmetry in terms of the Liouvillian generator~\cite{Kawabata2023,Gu2025,Albert2014}, in which coherent evolution and dissipation jointly determine the invariances of the dynamics.

This broader perspective has also transformed the role of dissipation in quantum science. Rather than being viewed solely as a source of decoherence, engineered system--environment coupling can be used to steer quantum dynamics toward desired states and phases. Early theoretical proposals established reservoir engineering as a route to many-body state preparation~\cite{Diehl2008,Verstraete2009}, followed by experimental demonstrations of dissipative entanglement stabilization in trapped ions~\cite{Lin2013} and superconducting circuits~\cite{Shankar2013}. Related approaches have been used to induce dissipative phase transitions~\cite{Diehl2010}, generate topological order~\cite{Diehl2011}, and protect quantum information~\cite{Harrington2022}.

Continuous measurement provides a particularly natural setting in which dissipation and symmetry become intertwined. Unlike projective measurements, which abruptly collapse the state, weak measurements continuously extract information while only gradually perturbing the system~\cite{Aharonov1988}. Such measurement backaction arises in superconducting qubits~\cite{Hacohen-Gourgy2020}, trapped ions~\cite{Pan2020}, nitrogen-vacancy centers, cavity-QED platforms, and quantum error-correction protocols. Quantum hardware provides a natural platform for simulating condensed-matter systems~\cite{Shen2026,Dalzell2025,Rosen2024}. It is therefore well suited for investigating how measurement backaction modifies dynamical symmetries. At the many-body level, repeated or continuous monitoring can generate qualitatively new dynamical behavior, including measurement-induced entanglement transitions~\cite{Li2018,Li2019,Skinner2019}. Recent work has also shown that inter-particle interactions can break tunneling symmetry that is preserved for noninteracting particles, even in the presence of an asymmetric barrier~\cite{Bilokon2025}. These developments raise a fundamental question: how does the choice of measured observable determine which dynamical symmetries survive and which are broken?
\vspace{-1.8pt}
\qquad In this work, we show that continuous weak measurements do not necessarily destroy dynamical symmetries. Instead, whether the symmetry is preserved or broken depends on the structure of the measurement operator. Using spin-1 tunneling as our primary setting, we identify the conditions under which measurement-induced dissipation leaves the tunneling symmetry intact and those under which it breaks it. We support this picture through analytical arguments and Lindblad simulations, and show how the same mechanism can be reproduced using quantum circuits based on Trotterized evolution and stochastic rotations.

\begin{figure*}
    \includegraphics[width=0.75\linewidth]{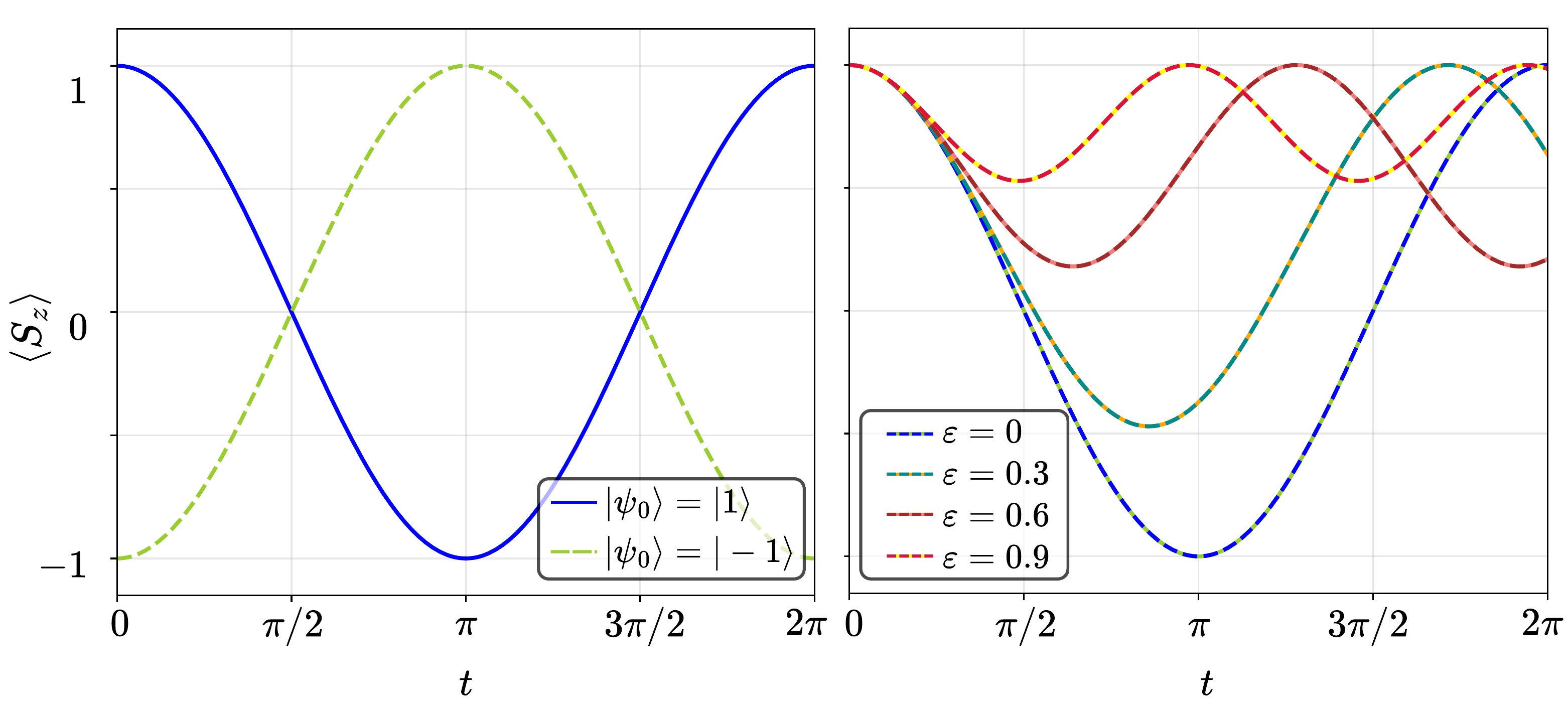}
    \caption{Expectation value $\langle S_z\rangle$ as a function of time for the Hamiltonian in Eq.~(\ref{Eq:Spin1_Hamiltonian}) with $D=-2$ and $\gamma=0.5$. The left panel shows the symmetric case ($\varepsilon=0$), characterized by complete tunneling between the states $\ket{+1}$ and $\ket{-1}$. The right panel corresponds to a finite asymmetry ($\varepsilon\neq0$), which suppresses complete population transfer while preserving the mirror symmetry of the dynamics. Solid and dashed curves correspond to the initial states $\ket{+1}$ and $\ket{-1}$, respectively, with the latter plotted as $-\langle S_z\rangle$. }
    \label{Fig:spin1_without_dissipation}
\end{figure*}

\section{Continuous Measurements and Open-System Dynamics}

A continuously monitored quantum system is naturally described as an open system. Its evolution follows the Lindblad master equation,
\begin{equation}
    \frac{d\rho}{dt} = \mathcal{L}[\rho] = -i[H,\rho]+\sum_k \mathcal{D}_{L_k}[\rho], 
    \label{Eq:Lindblad}
\end{equation}
where $H$ is the system Hamiltonian and
\begin{equation}
    \mathcal{D}_{L_k}[\rho] = L_k\rho L_k^\dagger-\frac{1}{2}\{L_k^\dagger L_k, \rho\}
\end{equation}
are the dissipative superoperators, with $L_k$ denoting the jump operators and $\rho$ is the density matrix. For a continuous measurement of a Hermitian observable $A$ with strength $p$, one may take
\begin{equation}\label{Eq:gen_dissip}
L = \sqrt{p}A,
\end{equation}
so that the dissipator reduces to a double commutator,
\begin{equation}
    \mathcal{D}_A[\rho] = -\frac{p}{2}\left[ A, [A,\rho]\right ].
    \label{Eq:dissipator}
\end{equation}

In an open system, symmetry must be formulated at the level of the full Liouvillian $\mathcal{L}$, not the 
Hamiltonian alone. A superoperator $\mathcal{S}[\rho] = U_{\mathcal{S}}\,\rho\,
U_{\mathcal{S}}^\dagger$ is a dynamical symmetry when $[\mathcal{L},\mathcal{S}]=0$. Whether a given dissipator preserves or breaks a symmetry of the Hamiltonian therefore depends on the interplay between the jump operator and the Hamiltonian: the jump operator must be analyzed in the context of the full Liouvillian rather than in isolation. In the following section, we show that the choice of measurement channel selectively preserves or breaks dynamical symmetry in the concrete setting.

\section{Measurement-Selective Symmetry Breaking in Spin-1 Tunneling}

\subsection{Spin-1 Hamiltonian and closed-system tunneling}
We first consider the coherent dynamics of a single spin-1 system governed by the Hamiltonian
\begin{equation}
    H = D S_z^2 + \gamma(S_x^2 - S_y^2) + \varepsilon S_z,
    \label{Eq:Spin1_Hamiltonian}
\end{equation}
where $S_\alpha$, with $\alpha\in\{x,y,z\}$, are spin-1 operators, $D$ is the axial constant that determines the magnetic anisotropy, and $\gamma$ is the transverse anisotropy parameter that drives coherent tunneling between $\ket{+1}$ and $\ket{-1}$ (see Ref.~\cite{Fernandez2009}). The last term introduces a longitudinal bias, with $\varepsilon$ controlling the energy splitting between the two tunneling states.

To characterize the relation between the two tunneling directions, we introduce the mirror operator
\begin{equation}
    \Pi = 
    \ket{+1}\!\bra{-1} +
    \ket{-1}\!\bra{+1} +
    \ket{0}\!\bra{0},
    \label{eq:mirror_operator}
\end{equation}
which exchanges the states $\ket{+1}$ and $\ket{-1}$ while leaving $\ket{0}$ unchanged. Its action on the spin operators is
\begin{equation}
    \Pi S_z \Pi^\dagger=-S_z,
    \qquad
    \Pi S_x \Pi^\dagger=S_x,
    \qquad
    \Pi S_y \Pi^\dagger=-S_y.
    \label{eq:mirror_spin_transformation}
\end{equation}
Consequently,
\begin{equation}
    \Pi H(\varepsilon)\Pi^\dagger = H(-\varepsilon).
    \label{eq:hamiltonian_covariance}
\end{equation}
For $\varepsilon=0$, the Hamiltonian is therefore exactly invariant under $\Pi$, and tunneling between $\ket{+1}$ and $\ket{-1}$ is perfectly symmetric. A finite bias breaks this symmetry, since $\Pi$ now maps $H(\varepsilon)$ to $H(-\varepsilon)$ rather than leaving the Hamiltonian invariant. See Appendix~\ref{apx:derivation} for more detail.

The resulting dynamics are shown in Fig.~\ref{Fig:spin1_without_dissipation}. For $\varepsilon=0$, the system undergoes complete coherent tunneling between $\ket{+1}$ and $\ket{-1}$ as shown in the left panel. When $\varepsilon\neq0$ (right panel), the energetic detuning suppresses complete population transfer. However, the dynamics initialized in $\ket{+1}$ and $\ket{-1}$ remain mirror-related. In the following subsection, we show that continuous monitoring may either preserve or destroy this symmetry, depending on the structure of the measured observable.

\begin{figure}[h!]
    \includegraphics[width=0.95\linewidth]{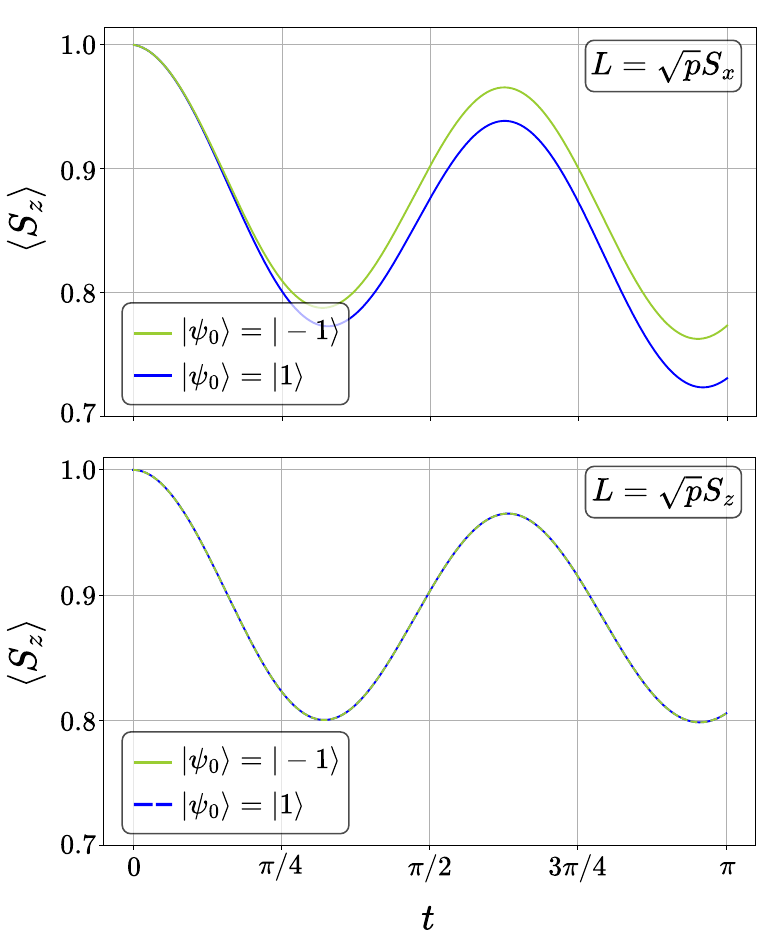}
    \caption{Expectation value $\langle S_z\rangle$ as a function of time under continuous weak measurement with $D = -2$, $\gamma = 0.5$, $\varepsilon = 1$, and $p = 0.05$. For initial state $|{-}1\rangle$, the observable plotted as $-\langle S_z \rangle$. Upper panel: jump operator $L = \sqrt{p_x}S_x$. Lower panel: jump operator $L = \sqrt{p_z}S_z$.}
    \label{Fig:symmetry_breaking}
\end{figure}

\subsection{Measurement-dependent tunneling dynamics}
So far, we have considered a closed quantum system. We now examine how continuous weak monitoring modifies the tunneling dynamics. The open-system dynamics are governed by the Lindblad equation with a single jump operator defined in Eq.\eqref{Eq:gen_dissip}, i.e.,
\begin{equation}
    L_\alpha=\sqrt{p_\alpha}\,S_\alpha,
    \qquad
    \alpha\in\{x,y,z\},
    \label{eq:spin_jump_operator}
\end{equation}
where $p_\alpha$ denotes the measurement strength. Although all three jump operators describe continuous measurements of a spin component, their effects on the tunneling dynamics are qualitatively different.

Figure~\ref{Fig:symmetry_breaking} compares the evolution under continuous monitoring of $S_x$ and $S_z$. In each case, the system is initialized either in $\ket{+1}$ or in $\ket{-1}$, and the latter dependence is plotted as $-\langle S_z(t)\rangle$ to facilitate comparison. Under $S_x$ monitoring, the two curves separate, meaning that the dissipator induces a clear asymmetry: the evolution starting from $|{+}1\rangle$ no longer mirrors that starting from $|{-}1\rangle$. By contrast, continuous monitoring of $S_z$ leaves the two trajectories indistinguishable. Despite the presence of both the bias term and the dissipator, the mirror symmetry of the tunneling is fully preserved. Whether the tunneling symmetry survives thus depends on the channel structure of the jump operator.

Since $S_x\ket{\pm 1}$ has nonzero overlap with $\ket{0}$, the dissipator with $L = \sqrt{p_x}\, S_x$ opens a decay channel through this intermediate level. The Hamiltonian bias $\varepsilon S_z$ in Eq.~\eqref{Eq:Spin1_Hamiltonian} splits the energies of $\ket{+1}$ and $\ket{-1}$, making the return transitions $\ket{0} \to \ket{+1}$ and $\ket{0} \to \ket{-1}$ energetically inequivalent and favoring one tunneling direction over the other. It is the combination of both ingredients that produces the observed tunneling asymmetry: the dissipator opening the $\ket{0}$ channel and the Hamiltonian bias weighting the return paths asymmetrically. Note that the jump operator $L = \sqrt{p_y}\, S_y$ produces analogous symmetry breaking, since $S_y$ also has nonzero overlap with the intermediate $\ket{0}$ state. By contrast, $S_z\ket{\pm 1} = \pm\ket{\pm 1}$ never populates $\ket{0}$, and the resulting dissipator amounts to pure dephasing within the tunneling subspace, preserving the mirror symmetry for any $\varepsilon$.

More generally, a dissipator breaks the tunneling symmetry only if the corresponding jump operator has nonzero $\Delta m_s = \pm 1$ matrix elements, i.e., it couples the tunneling subspace to the intermediate $|0\rangle$ state. The bias $\varepsilon$ then converts this coupling into a measurable asymmetry.

\begin{figure*}[t]
    \includegraphics[width=0.9\textwidth]{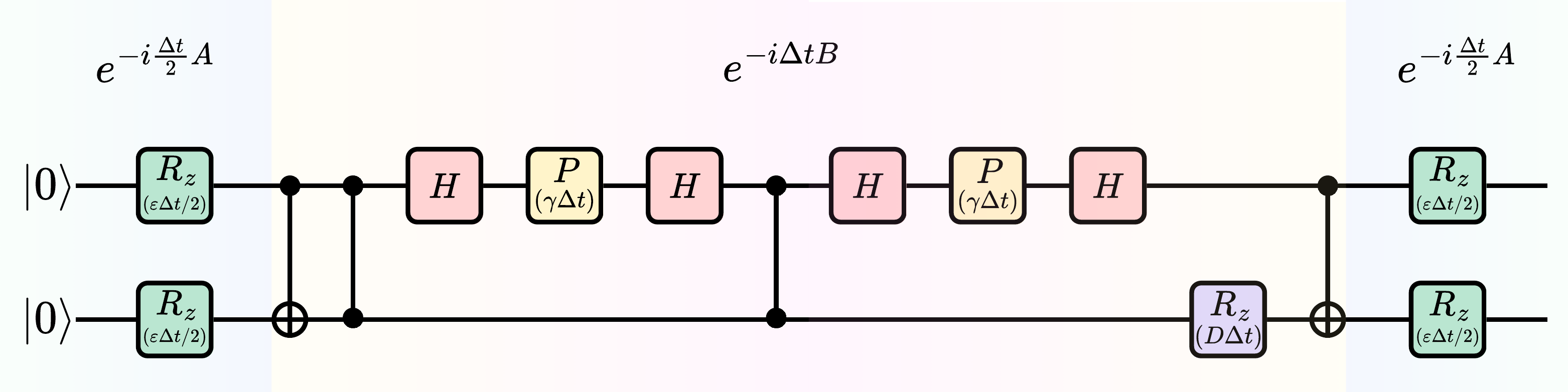}
    \caption{Quantum circuit for one Trotter step used to simulate the dynamics generated by the Hamiltonian in Eq.~(\ref{Eq:Spin1_Hamiltonian}). The full-time evolution is obtained by repeating this circuit with a time step $\Delta t$. The Hamiltonian is decomposed into the asymmetric part, $A=\varepsilon S_z$, and the symmetric part, $B=D S_z^2+\gamma(S_x^2-S_y^2)$. The gates $H$, $P$, and $R_z$ represent the Hadamard, phase, and $Z$-rotation gates, respectively. The circuit also employs controlled-NOT (CNOT) and controlled-$Z$ (CZ) gates to implement two-qubit interactions.}
    \label{Fig:trotter_step_circuit}
\end{figure*}

\subsection{Simulation using Quantum-Circuits}
A spin-1 system can be realized using two spin-$1/2$ particles (see Ref.~\cite{Sinha2003}). In this representation, the spin-1 operators are expressed as

\begin{equation}
S_\alpha = \frac{1}{2}\left(\sigma_\alpha^1+\sigma_\alpha^2\right),
\end{equation}

where $\sigma_\alpha^i$ are Pauli operators and $i=1,2$. Using this representation, the Hamiltonian in Eq.~\eqref{Eq:Spin1_Hamiltonian} can be rewritten as
\begin{equation}
H = \frac{D}{2}\left(1+\sigma_z^1\sigma_z^2\right)
+ \frac{\gamma}{2}\left(\sigma_x^1\sigma_x^2-\sigma_y^1\sigma_y^2\right)
+ \frac{\varepsilon}{2}\left(\sigma_z^1+\sigma_z^2\right).
\end{equation}
For $\varepsilon=0$, the Hamiltonian is symmetric and all its terms commute. Consequently, the evolution operator can be factorized as

\begin{equation}
U(t)=e^{-itH}
=e^{-i\frac{D}{2}t}
e^{-i\frac{D}{2}\sigma_z^1\sigma_z^2 t}
e^{-i\frac{\gamma}{2}\sigma_x^1\sigma_x^2 t}
e^{i\frac{\gamma}{2}\sigma_y^1\sigma_y^2 t}.
\end{equation}

Such an evolution can be implemented using a quantum circuit (see Ref.~\cite{Gnatenko2023}). In the absence of the asymmetry term, the system dynamics can be simulated by modifying only the circuit parameters, while keeping the circuit depth unchanged.

The asymmetric term $\varepsilon S_z$, however, does not commute with the remaining terms of the Hamiltonian. In this case, the system dynamics can be simulated using a Trotter decomposition of the unitary evolution operator (see Refs.~\cite{Hatano2005,Berry2007}). We employ the second-order Trotter--Suzuki formula,
\begin{equation}
e^{-it(A+B)}
\approx
e^{-i\frac{t}{2}A}
e^{-itB}
e^{-i\frac{t}{2}A},
\end{equation}
where $A = \varepsilon S_z$ is the bias term and $B = D S_z^2 + \gamma(S_x^2 - S_y^2)$ contains the remaining symmetric terms.  One Trotter step of the corresponding quantum circuit is shown in Fig.~\ref{Fig:trotter_step_circuit}. As the Trotter time step is decreased, the approximation becomes increasingly accurate. The error of the second-order Trotter--Suzuki decomposition scales as $O(\Delta t^3)$ per step. 

\begin{figure*}[t]
    \includegraphics[width=0.85\textwidth]{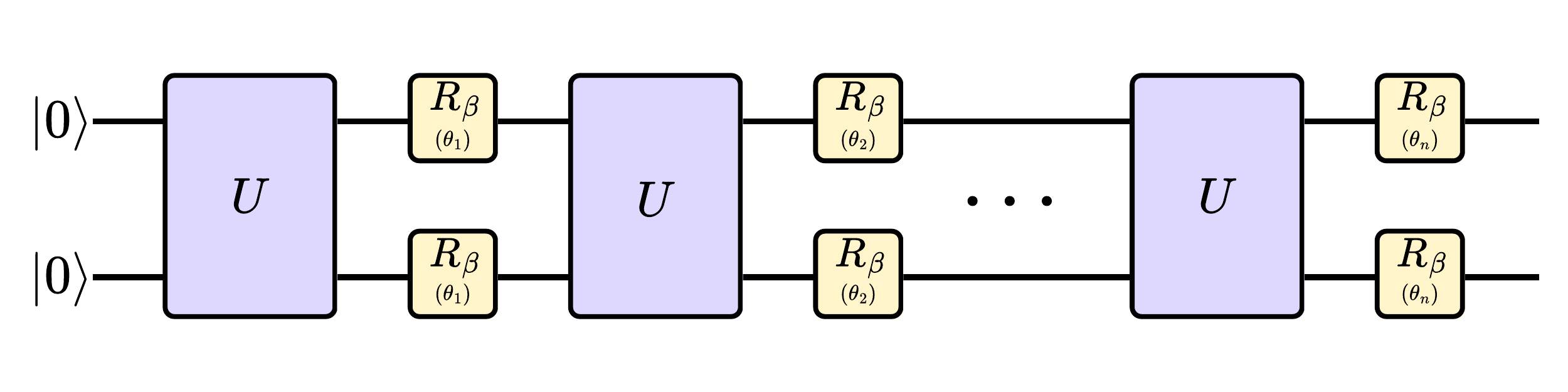}
    \caption{Full quantum circuit with dissipation. Each block $U$ represents one Trotter step as shown in Fig.~\ref{Fig:trotter_steps}. Dissipation is implemented by applying $R_\beta(\theta_i)$ $(\beta=x,z)$ rotations to both qubits with the same angle $\theta_i$ between successive Trotter steps. All angles $\theta_i$ are sampled uniformly in range $\left [-\sqrt{3p_\beta\Delta t}, +\sqrt{3p_\beta\Delta t}\right ]$.}
    \label{Fig:spin1_full_circuit}
\end{figure*}

The dissipative evolution of the system can be realized using averaging over quantum trajectories~\cite{Landi2024}. Each trajectory consists of the Trotterized unitary evolution with $R_\beta(\theta)$ rotations ($\beta=x, z$) applied to both qubits between successive Trotter steps
\begin{equation}
    R_\beta(\theta)\otimes R_\beta(\theta)=e^{-i\frac{\theta}{2}\sigma_\beta^{1}}e^{-i\frac{\theta}{2}\sigma_\beta^{2}}=e^{-i\theta S_\beta}.
\end{equation}
A single stochastic time step transforms the density matrix as
\begin{equation}
    \rho^\prime = U_\theta\rho U_\theta^\dagger=e^{-i\theta S_\beta}\rho e^{i\theta S_\beta}.
\end{equation}
For small $\theta$, appropriate to the continuous weak measurement regime, expanding to second order gives
\begin{equation}
    e^{-i\theta S_\beta}\rho e^{i\theta S_\beta} = \rho -i\theta [S_\beta, \rho]-\frac{\theta^2}{2}[S_\beta,[S_\beta,\rho]] +O(\theta^3).
\end{equation}
Averaging over $\theta$ drawn uniformly from $[-\theta_0, \theta_0]$, the first-order term vanishes since $\langle\theta\rangle=0$, yielding
\begin{equation}
    \mathrm{E}[\rho^\prime] = \rho -\frac{\langle\theta^2\rangle}{2}[S_\beta,[S_\beta,\rho]].
\end{equation}
On the other hand, a single dissipative time step with dissipator (\ref{Eq:dissipator}) is 
\begin{equation}
    \rho(t+\Delta t) = \rho -\frac{p_\beta\Delta t}{2}[S_\beta,[S_\beta,\rho]].
\end{equation}
Then, to describe the dissipative evolution with averaging over stochastic dynamics, we get $\langle\theta^2\rangle=p_\beta\Delta t$. For uniform sampling $\langle\theta^2\rangle=\theta_0^2/3$, giving the rotation range $\theta_0 = \sqrt{3p_\beta\Delta t}$.

\subsection{Quantum Simulation Results}
\begin{figure}[h!]
    \includegraphics[width=0.45\textwidth]{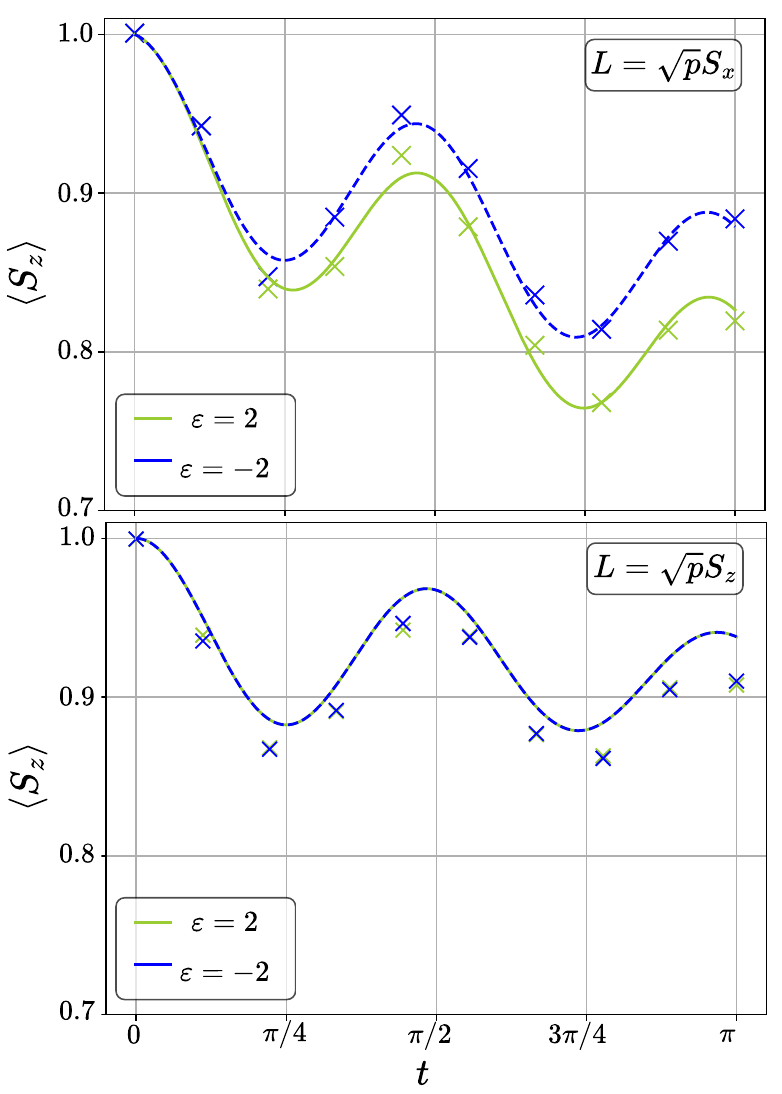}
    \caption{Time evolution of the spin-1 system for the $S_x$ and $S_z$ dissipative channels. The system parameters are $D=-2$, $\gamma=0.2$, and $p_\beta=0.1$, where $\beta=x,z$. The green curves correspond to $\varepsilon=2$, while the blue curves correspond to $\varepsilon=-2$. The solid curves show the solution of the Lindblad equation. The $\times$ markers represent the results of the quantum circuit simulation performed on the noisy simulator of \textrm{ibm\_marrakesh}. The Operator Decoherence Renormalization (ODR) error mitigation technique is applied to reduce the effect of noise. Each data point is obtained by averaging over 100 quantum circuits with independently sampled random rotation angles. For each circuit, 1000 measurement shots are performed.}
    \label{Fig:result_simulation}
\end{figure}

To simulate the dynamics of a spin-1 system with the Hamiltonian (\ref{Eq:Spin1_Hamiltonian}) and dissipators $S_x$ and $S_z$, we use the full quantum circuit shown in Fig.~\ref{Fig:spin1_full_circuit}. At each Trotter step, $\theta_i$ is sampled uniformly from the interval $\left[-\sqrt{3p_\beta\Delta t}, +\sqrt{3p_\beta\Delta t}\right]$. The simulation results are presented in Fig.~\ref{Fig:result_simulation}. The time evolution is obtained by averaging over $100$ quantum circuits. For each circuit, we perform $1000$ measurement shots using the noisy simulator of \textrm{ibm\_marrakesh}. Since the simulations are carried out on a noisy quantum device simulator, the results are not ideal. To reduce the impact of noise, we employ the Operator Decoherence Renormalization (ODR) error mitigation method~\cite{Urbanek2021}.

Despite the presence of simulation errors, we observe that the dynamical symmetry is broken in the case of $S_x$ dissipation. The upper panel of Fig.~\ref{Fig:result_simulation} shows the evolution for the two cases, $\varepsilon=2$ and $\varepsilon=-2$. The asymmetry is the same as that discussed in Appendix~\ref{apx:derivation}: the evolution starting from $|{+}1\rangle$ no longer mirrors that starting from $|{-}1\rangle$. The lower panel shows the results for the $S_z$ dissipator, which preserves the dynamical symmetry.

\section*{Conclusion}

We have shown that continuous weak measurements can selectively break dynamical symmetries that persist in the closed-system dynamics. Using spin-1 tunneling as a minimal example, we found that measurements of $S_x$ and $S_y$ destroy the tunneling symmetry between the $\ket{+1}$ and $\ket{-1}$ states, whereas measurement of $S_z$ preserves it. The distinction originates from the available measurement-induced channels: $S_x$ and $S_y$ couple the tunneling states to the intermediate $\lvert 0\rangle$ level, while $S_z$ produces dephasing without leaving the $\{\lvert +1\rangle,\lvert -1\rangle\}$ subspace. In the presence of the longitudinal bias, these additional channels distinguish the two tunneling directions and produce an observable dynamical asymmetry.

We further showed that the same measurement-induced evolution can be implemented in a two-qubit quantum simulation using Trotterized unitary dynamics and stochastic collective rotations, providing a direct route toward realizing the effect on quantum hardware. These results establish continuous measurement as a selective control mechanism for preserving or removing dynamical relations, with possible extensions to interacting spin systems and many-body quantum simulators.

\acknowledgments
P.K. acknowledges financial support from FCT, Portugal, through the individual grant 2024.01084.BD. This work acknowledges financial support by CeFEMA, Centre of Physics and Engineering of Advanced Materials, under contracts UID/04540/2025, UID/PRR/04540/2025 and UID/PRR2/04540/2025 with the Portuguese Agência para a Investigação e Inovação AI2, doi https://doi.org/10.54499/UID/PRR/04540/2025, https://doi.org/10.54499/UID/PRR/04540/2025 and https://doi.org/10.54499/UID/PRR2/04540/2025.
P.K. also acknowledges support from the American Physical Society through the DIRAC Program.

E.B. and D.I.B. were supported by Army Research Office (ARO) (grant W911NF-23-1-0288; program manager Dr.~James Joseph).
 V.B. is supported by NASA EPSCoR and/or the Board of Regents Support Fund.  The views and conclusions contained in this document are those of the authors and should not be interpreted as representing the official policies, either expressed or implied, of ARO, NASA, the Board of Regents, or the U.S. Government. The U.S. Government is authorized to reproduce and distribute reprints for Government purposes notwithstanding any copyright notation herein.

\appendix
\section{Derivation}\label{apx:derivation}
We study dynamical symmetry breaking, meaning that dynamics starting from different initial states has symmetry in the non-dissipative case. By introducing dissipation, one can either break or preserve this symmetry. To make the distinction precise, we introduce the mirror operator
\begin{equation}
\Pi
=
\ket{+1}\bra{-1}
+
\ket{-1}\bra{+1}
+
\ket{0}\bra{0},
\end{equation}
which exchanges the two tunneling states while leaving the intermediate state unchanged. Its action on the spin-1 operators is
\begin{equation}
\Pi S_z\Pi^\dagger=-S_z,\qquad
\Pi S_x\Pi^\dagger=S_x,\qquad
\Pi S_y\Pi^\dagger=-S_y.
\end{equation}

The corresponding transformation of density matrices is the superoperator
\begin{equation}
\mathcal P(\rho)=\Pi\rho\Pi^\dagger.
\end{equation}

For a Lindblad equation
\begin{equation}
\dot{\rho}
=
\mathcal L_\epsilon(\rho)
=
-i[H_0+\epsilon S_z,\rho]
+
\mathcal D_L(\rho),
\end{equation}
with
\begin{equation}
\mathcal D_L(\rho)
=
L\rho L^\dagger
-\frac12\{L^\dagger L,\rho\},
\end{equation}
the dissipative and Hamiltonian contributions should be analyzed separately.

If the unbiased Hamiltonian is mirror invariant,
\begin{equation}
\Pi H_0\Pi^\dagger=H_0,
\end{equation}
then the bias transforms as
\begin{equation}
\Pi(\epsilon S_z)\Pi^\dagger=-\epsilon S_z.
\end{equation}

Consequently, the Hamiltonian does not commute with the mirror transformation at fixed nonzero $\epsilon$. Instead, it satisfies the covariance relation
\begin{equation}
\mathcal P H(\epsilon)
=
H(-\epsilon)\,\mathcal P.
\end{equation}

This relation implies a symmetry between the evolution of the states $\vert 1\rangle$ and $-\vert -1\rangle$, analogous to the symmetry between the dynamics generated by the Hamiltonians $H(\varepsilon)$ and $H(-\varepsilon)$.

The distinction between dissipators that preserve or break the mirror symmetry can be understood from their commutators with the Hamiltonian. Using
\begin{equation}
[S_z,S_x]=iS_y,\qquad
[S_z,S_y]=-iS_x,
\end{equation}
one can calculate the commutator between longitudinal bias term of the Hamiltonian and the corresponding jump operator, i.e.,
\begin{equation}
[H,L_z]=[\epsilon S_z,L_z]=0,
\end{equation}
whereas
\begin{equation}
[H,L_x]=[\epsilon S_z,L_x]
=
i\epsilon\sqrt{p_x}\,S_y,
\end{equation}
\begin{equation}
[H,L_y]=[\epsilon S_z,L_y]
=
-i\epsilon\sqrt{p_y}\,S_x.
\end{equation}
This shows that $S_x$ and $S_y$ dissipators break the dynamical symmetry while $S_z$ preserves it.

\section{Liouvillian zero modes and dynamical symmetry}

\begin{figure}[t]
    \centering
    \includegraphics[width=0.9\columnwidth]{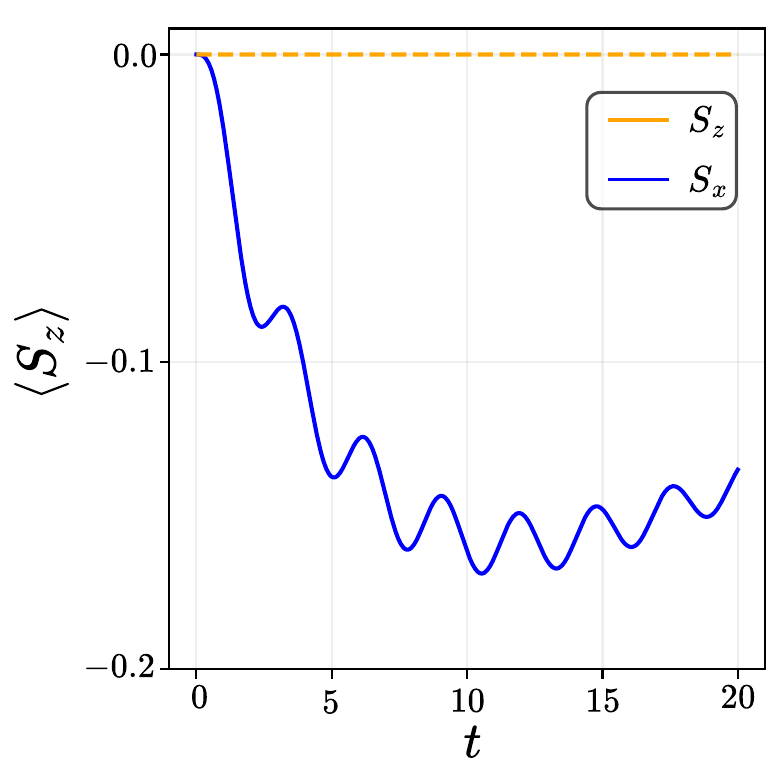}
    \caption{
        Time evolution of $\langle S_z(t)\rangle$ (orange curve) for the initial
        state $|0\rangle$. For measurement of $S_z$, the intermediate
        state is dynamically disconnected and
        $\langle S_z(t)\rangle_0=0$. For measurement of $S_x$  (blue curve), the
        intermediate state is coupled to $|\pm1\rangle$ and develops
        a nonzero transient magnetization at finite bias. 
    }
    \label{fig:zero_state_dynamics}
\end{figure}

The different dynamical behavior produced by measurements of $S_z$ and $S_{x,y}$ can be understood from the zero modes of the corresponding Liouvillian. We consider
\begin{equation}
    \mathcal{L}_{\alpha}(\rho)
    =
    -i[H,\rho]
    +
    p_{\alpha}
    \left(
        S_{\alpha}\rho S_{\alpha}
        -
        \frac{1}{2}\{S_{\alpha}^{2},\rho\}
    \right),
    \label{eq:liouvillian_channels}
\end{equation}
where $\alpha=x,y,z,$, and 
\begin{equation}
    H
    =
    D S_z^2
    +
    \gamma(S_x^2-S_y^2)
    +
    \epsilon S_z.
    \label{eq:spin_one_hamiltonian_zero_modes}
\end{equation}
A right zero mode $R_j$ of the Liouvillian satisfies
\begin{equation}
    \mathcal{L}_{\alpha}(R_j)=0
\end{equation}
and represents a stationary operator. The number of
zero eigenvalues of $\mathcal{L}_{\alpha}$ therefore determines the number of linearly independent stationary modes.

Introduce the projectors
\begin{equation}
    P_0=|0\rangle\langle 0|,
    \qquad
    P_{\pm}
    =
    |+1\rangle\langle+1|
    +
    |-1\rangle\langle-1|
    =
    I-P_0.
    \label{eq:sector_projectors}
\end{equation}

For the $S_z$ measurement channel, both the Hamiltonian and the jump
operator preserve the two subspaces,
\begin{equation}
    [H,P_0]=0,
    \qquad
    [S_z,P_0]=0.
    \label{eq:sz_disconnected_sector}
\end{equation}
Consequently, the state $|0\rangle$ is dynamically disconnected from
the tunneling subspace spanned by $\{|+1\rangle,|-1\rangle\}$. In the
Heisenberg picture,
\begin{equation}
    \mathcal{L}_z^\dagger(P_0)=0,
    \label{eq:p0_conserved}
\end{equation}
so that the population of the intermediate state is conserved:
\begin{equation}
    \frac{d}{dt}\operatorname{Tr}[P_0\rho(t)]=0.
\end{equation}

The $S_z$ Liouvillian therefore contain two independent zero modes, which may be chosen as
\begin{equation}
    \rho_{\mathrm{ss}}^{(0)}
    =
    |0\rangle\langle0|,
    \qquad
    \rho_{\mathrm{ss}}^{(\pm)}
    =
    \frac{1}{2}
    \left(
        |+1\rangle\langle+1|
        +
        |-1\rangle\langle-1|
    \right).
    \label{eq:sz_stationary_modes}
\end{equation}

The physical stationary states form the family
\begin{equation}
    \rho_{\mathrm{ss}}(q)
    =
    q\,\rho_{\mathrm{ss}}^{(0)}
    +
    (1-q)\rho_{\mathrm{ss}}^{(\pm)},
    \qquad
    0\leq q\leq1,
    \label{eq:sz_stationary_family}
\end{equation}
where $q=\langle0|\rho(0)|0\rangle$ is fixed by the initial condition.

Because the dynamics starting from $|\pm1\rangle$ remains confined to
the effective two-level tunneling subspace, the transition dynamics
retains the same-bias reciprocity
\begin{equation}
    \langle S_z(t)\rangle_{+}
    =
    -\langle S_z(t)\rangle_{-},
    \label{eq:sz_reciprocity}
\end{equation}
where
\begin{equation}
    \langle S_z(t)\rangle_m
    =
    \operatorname{Tr}
    \left[
        S_z e^{t\mathcal{L}_{\alpha}}
        (|m\rangle\langle m|)
    \right].
\end{equation}

In case of $S_x$ or $S_y$ measurements, these
operators have nonzero matrix elements between the intermediate state
and the tunneling states:
\begin{equation}
    \langle0|S_x|\pm1\rangle\neq0,
    \qquad
    \langle0|S_y|\pm1\rangle\neq0.
\end{equation}
Equivalently,
\begin{equation}
    [S_x,P_0]\neq0,
    \qquad
    [S_y,P_0]\neq0.
    \label{eq:transverse_channel_mixing}
\end{equation}
The transverse measurement channels therefore induce transitions
\begin{equation}
    |+1\rangle
    \longleftrightarrow
    |0\rangle
    \longleftrightarrow
    |-1\rangle,
\end{equation}
and the population of $|0\rangle$ is no longer conserved:
\begin{equation}
    \mathcal{L}_{x,y}^{\dagger}(P_0)\neq0.
\end{equation}

As a result, for generic nonzero parameters there is one zero eigenvalue of $\mathcal{L}_{x,y}$.

The evolution starting from $|0\rangle$ therefore provides a direct
diagnostic of this effect. As shown in
Fig.~\ref{fig:zero_state_dynamics}, $S_z$ measurement leaves
$\langle S_z(t)\rangle_0$ identically zero, whereas an $S_x$
measurement at finite bias generates a nonzero transient value. 

The number of zero modes and the symmetrical transition dynamics are
not generally equivalent. In the present model, however, they are
controlled by the same physical mechanism: whether the dissipative
channel connects the tunneling subspace to the intermediate
$|0\rangle$ state. Opening this channel removes the additional
stationary mode and, in combination with the finite bias, enables
transient nonreciprocal tunneling. 

\nocite{apsrev41Control} 
\bibliographystyle{apsrev4-1}
\bibliography{bibliography,1}
\end{document}